\documentclass[12pt]{article}
\usepackage{epsfig}
\begin{document}

\def\bib#1{[{\ref{#1}}]}
\title{\bf Radon transform of Wheeler-De Witt equation
 and tomography of quantum states of the universe}
\author{{V.~I.~Man'ko${ }^{1,2}$, G.~Marmo${ }^{1,2}$ and C.~Stornaiolo ${ }^{1,2}$}\\
{\em $~^{1}$ Istituto Nazionale di Fisica Nucleare,}
{\em Sezione di Napoli,}\\
 {\em  Complesso Universitario di Monte S. Angelo}
 {\em Edificio N' }\\ {\em via Cinthia, 45 -- 80126 Napoli}\\
{\em $~^{2}$ Dipartimento di Scienze
Fisiche,}\\ {\em Universit\`{a} ``Federico II'' di Napoli,}\\
 {\em  Complesso Universitario di Monte S. Angelo}
 {\em Edificio N'  }\\ {\em via Cinthia, 45 -- 80126  Napoli}\\ }
\date{ }
\maketitle

\begin{abstract}
The notion of standard positive probability distribution function
(tomogram) which describes the quantum state of universe
alternatively to wave function or to density matrix is introduced.
Connection of the tomographic probability distribution with the
Wigner function of the universe and with the star-product
(deformation) quantization procedure is established.

Using the Radon transform the Wheeler-De Witt generic equation for
the probability function is written in tomographic form.  Some
examples of the Wheeler-DeWitt equation in the minisuperspace are
elaborated explicitly for a homogeneous isotropic cosmological
models. Some interpretational aspects of the probability
description of the quantum state are discussed.
\end{abstract}

\vskip 0.5truecm
\section{Introduction}

Recently in conventional quantum mechanics the Radon \bib{radon}
transform of the von Neumann density operator
\bib{vonneumann22}\bib{landau} considered in the form of Wigner
function \bib{wigner32} was recognized to give the tomographic
probability (called tomographic map or tomogram) appropriate to
reconstruct quantum states
\bib{vogrisken}\bib{raymer}\bib{bertr}. The slightly modified
Radon transform of density matrix with additional scaling
transform was suggested \bib{mancini95} in which the problem of
the singularity of the Radon transform using only a rotation
parameter was smoothed.

The possibility to implement the tomographic probability to define
the quantum state in terms of conventional probability was
suggested in \bib{mancinipl}. It was pointed out that there exists
a representation in quantum mechanics in which any quantum state
can be described by a standard positive probability distribution.
In \bib{marmosud} it was shown how superposition principle
(quantum interference) is described using only positive
probabilities. The properties of the tomographic map and its
relation to the Heisenberg-Weyl group, SU(2)-group and to
star-product quantization were discussed in
\bib{marmops}.

It was understood \bib{marmops} that the tomographic map is
closely related to the star-product quantization procedure which
provides Moyal equation for quantum evolution in the phase-space
representation of quantum states \bib{moyal49}

On the other hand quantum cosmology uses as basic notion the wave
function of the universe (it would be better to say the wave
functional) which depends on the metric and the material fields
\bib{hawking1}. This wave  functional obeys the Wheeler-De Witt
equation \bib{whdw} which is a  generalisation of the
Schr\"odinger equation for the wave function. The properties of
the Wheeler-De Witt description of the universe are the subject of
an intensive discussion \bib{hh}\bib{hawking2}\bib{dnp}\bib{hl}
due to the importance of this approach in quantum cosmology (see
review in
\bib{rubakov}).

Interpretation of the wave function of the universe contains the
same problems as the interpretation of the wave function of finite
quantum systems\bib{parentani}. The notion of density matrix of
the universe is also used to describe the state of the universe
(see for example\bib{dnp}). The Weyl-Wigner representation of the
density matrix and the corresponding deformation  quantization
procedure was used  within the context of cosmological problems in
\bib{anton} and \bib{kodama}.

The general study of deformation quantization in quantum gravity
is a highly non trivial procedure \bib{anton}. In our paper we use
a particular deformation procedure related to the tomographic star
product formalism \bib{marmops}.

The aim of our work is to introduce the tomographic probability
function which describes the state of the universe and contains
the same information on the universe state as the density matrix
does. To reach this aim we apply the modified Radon transform for
the density matrix of the universe, which is using the extension
of the modified Radon transform of \bib{mancini95} for one degree
of freedom. The Radon transform of the density matrix can be cast
into the form of the transform of the wave-function
\bib{mendespl}. We will discuss the extension of the functional
Radon transform both in the form of the transform of the wave
functional of the universe and in the form of the transform of the
density matrix functional. Our procedure is an heuristic one and
the rigorous mathematics of functional measures and of convergence
of integral functionals needs further investigations, see for
example
 \bib{asht} \bib{baez}.

Our goal is to obtain the Wheeler-De Witt  equation in tomographic
(or probability) representation written for the tomogram of the
universe. We consider also the simplest case of Wheeler-De Witt
equation for the wave functional in which all the variety of
metrics is reduced to the variety of radial time dependence. In
this case the Wheeler-De Witt equation takes the form of
Schr\"{o}dinger-like equation for one degree of freedom. We will
reobtain this equation in the form of the equation for tomographic
probability.

The paper is organized as follows. In the next section 2 we review
the star product procedure (or the deformation procedure) in a
general form. In section 3 we present the tomographic approach in
the phase space. In section 4 the functional Radon transform is
discussed. In section 5  examples of tomographic representation of
the Wheeler-DeWitt equation for one degree of freedom are given.
Using the extension of the tomographic functional of a scalar
field \bib{rosapl} the notion of probability functional of the
universe is introduced in Appendix.

Also the generic Wheeler-DeWitt equation is written in the form of
von Neumann like equation for density matrix functional and by
means of the functional Radon transform it is rewritten in
tomographic form in Appendix. Perspectives of the probability
description of the universe state are discussed in the
conclusions, section 6.

\section{Star-product deformation quantization}
In this section we review the quantization procedure. This
approach uses a deformation procedure \bib{anton}. Another
explanation of the procedure is introducing and using a
star-product of the operator symbols \bib{fronsdal}. Below we
follow the presentation of the star-product as given in
\bib{marmops}. Let us consider a Hilbert space $\mathcal{H}$ and
a set of operators acting in the Hilbert space. The state of the
universe can be associated either with a vector in the Hilbert
space or with a density operator $\hat{\rho}$ which is a
nonnegative Hermitian operator. Let us consider an operator
$\hat{A}$. Let us suppose also that there exists a set of
operators $\hat{U}(\vec{x})$ where
$\vec{x}=(x_{1},x_{2}...,x_{N})$ such that the function (called
the symbol of the operator $\hat{A}$)
\begin{equation}\label{symbA}
    f_{\hat{A}}(\vec{x})=Tr(\hat{A}\hat{U}(\vec{x}))
\end{equation}
defines the operator completely. It means that there exists a dual
set of operators $\hat{{\mathcal D}}(\vec{x})$ such that one has
the relation
\begin{equation}\label{opA}
    \hat{A}=\int  f_{\hat{A}}(\vec{x})\hat{{\mathcal{D}}}(\vec{x})d\vec{x}.
\end{equation}
 If there exist such operator families $\hat{U}(\vec{x})$ and $\hat{{\mathcal
 D}}(\vec{x})$, one can introduce the star-product of symbols defined by the relation
\begin{equation}\label{starproduct}
f_{\hat{A}\hat{B}}(\vec{x})= f_{\hat{A}}(\vec{x})*
f_{\hat{B}}(\vec{x}):=Tr(\hat{A}\hat{B}\hat{U}(\vec{x})).
\end{equation}
 In view of the associativity of the operator product the star-product
 is also associative, i.e.
\begin{equation}\label{associat}
    f_{\hat{A}}(\vec{x})* (f_{\hat{B}}(\vec{x})*f_{\hat{C}}(\vec{x}))=
    (f_{\hat{A}}(\vec{x})* f_{\hat{B}}(\vec{x}))*f_{\hat{C}}(\vec{x}).
\end{equation}
In our article we shall discuss two types of symbols associated
with operators. The first type is called the Weyl symbol. For a
state density operator Weyl symbol is the Wigner function. The
Weyl symbol $W_{\hat{A}}(q,p)$ of an operator $\hat{A}$ is defined
by the following families of operators

 \begin{equation}\label{1}
\hat{U}(\vec{x})=\hat{U}(x_{1},x_{2})
\end{equation}

 and

 \begin{equation}\label{2}
\hat{{\mathcal D}}(\vec{x})=\hat{{\mathcal D}}(x_{1},x_{2})
\end{equation}
for which we assume  $x_{1}=q/ \sqrt{2}$,
 $x_{2}=p/\sqrt{ {2}}$ (we consider a quantum system with one degree of
 freedom). Thus we introduce the two family of operators
 \begin{equation}\label{U}
    \hat{U}(q,p)=2\hat{\mathcal{D}}(\alpha)(-1)^{a^{\dag}
    a}\hat{\mathcal{D}}(-\alpha), \,\,\,\,\, \alpha=\frac{1}{\sqrt{2}}\left({q}+
    i{p}\right),
\end{equation}
\begin{equation}\label{D}
    \hat{\mathcal{D}}(q,p)=\frac{1}{\pi}\hat{\mathcal{D}}(\alpha)(-1)^{a^{\dag}
    a}\hat{\mathcal{D}}(-\alpha).
\end{equation}
Here the operators $a^{\dag}$ and $a$ are bosonic creation and
annihilation operators
 \begin{equation}\label{annih}
    a=\frac{1}{\sqrt{2}}\left(\hat{q}+ i\hat{p}\right).
\end{equation}
In position representation the operators $\hat{q}$ and $\hat{p}$
are given by the standard relation
 \begin{equation}\label{standard}
   \hat{ q}\psi(x)=x\psi(x), \ \ \ \ \ \ \ \ \
   \hat{p}\psi(x)=-i\frac{\partial\psi}{\partial x},
\end{equation}
$\hbar =1$.

Note that the introduced operators have two aspects, one related
to the linear transformations in the coordinate space, the other
has to do with the unitary representations of the translation
group.

In (\ref{U}) and (\ref{D}) the operator $\hat{{\mathcal
D}}(\alpha)$, where $\alpha$ is the complex number defined in
(\ref{U}), is a unitary displacement operator
\begin{equation}\label{dalfa}
     \hat{{\mathcal D}}(\alpha)= \exp(\alpha a^{\dag} -\alpha^{*} a).
\end{equation}
%

The operator $(-1)^{a^{\dag} a}$ is the parity operator. Thus the
Weyl symbol of the operator $\hat{A}$ is defined by the relation
\begin{equation}\label{weylsymb}
     W_{\hat{A}}=2Tr(\hat{A}\hat{\mathcal{D}}(\alpha)(-1)^{a^{\dag} a}
     \hat{\mathcal{D}}(-\alpha).
\end{equation}
If $\hat{A}$ is a density operator $\hat{\rho}$ defining a state
of a quantum system (the state of the universe) the relation
(\ref{weylsymb}) provides the Wigner function of the state. In our
paper we will study another symbol
${\mathcal{W}}_{\hat{A}}(X,\mu,\nu)$ of the operator $\hat{A}$
called the tomographic symbol.

The tomographic symbol is defined by means of the pair of families
of operators $\hat{U}(\vec{x})$ and $\hat{{\mathcal
 D}}(\vec{x})$ where $\vec{x}=(X,\mu ,\nu)$
and $X$, $\mu $, $\nu$ are real numbers. The operators are given
by the formulae
\begin{equation}\label{utomo}
\hat{U}(X,\mu ,\nu)= \delta(X-\mu\hat{q}-\nu\hat{p})
\end{equation}
\begin{equation}\label{dtomo}
    \hat{{\mathcal
 D}}(X,\mu ,\nu)=\frac{1}{2\pi}e^{iX}e^{i(\mu\hat{q}+\nu\hat{p})}.
\end{equation}
Thus the symbol of an operator $\hat{A}$, called the tomogram, is
given by the relation
\begin{equation}\label{wtomo}
   { \mathcal W}_{\hat{A}}(X,\mu ,\nu)=Tr(\hat{A}\delta(X-\mu\hat{q}-\nu\hat{p})).
\end{equation}
The inverse relation reads
\begin{equation}\label{inversetomo}
     \hat{A}=\frac{1}{2\pi}\int{\mathcal W}_{\hat{A}}(X,\mu ,\nu)
     e^{iX+i\mu\hat{q}+i\nu\hat{p}}dX d\mu d\nu.
\end{equation}
The star-product of two tomograms is defined using the kernel
$$
{ \mathcal W}_{\hat{A}}(X,\mu ,\nu)*{ \mathcal W}_{\hat{B}}(X,\mu
,\nu)= \int dX_{1} d\mu_{1} d\nu_{1} dX_{2} d\mu_{2} d\nu_{2}
{\mathcal W}(X_{1},\mu_{1} ,\nu_{1})
$$
\begin{equation}\label{startomo}
\times{\mathcal W} (X_{2},\mu_{2} ,\nu_{2})K(X_{1},\mu_{1}
,\nu_{1},X_{2},\mu_{2} ,\nu_{2},X,\mu ,\nu)
\end{equation}
Here the kernel is given by
\begin{equation}\label{tracekernel}
K(X_{1},\mu_{1},\nu_{1},X_{2},\mu_{2} ,\nu_{2},X,\mu ,\nu)=
Tr\left\{ \hat{{\mathcal
 D}}(X_{1},\mu_{1},\nu_{1})\hat{{\mathcal
 D}}(X_{2},\mu_{2},\nu_{2})\hat{U}(X,\mu ,\nu)\right\}.
\end{equation}
This trace can be calculated and the tomographic kernel reads
\bib{marmops}
$$
K(X_{1},\mu_{1},\nu_{1},X_{2},\mu_{2} ,\nu_{2},X,\mu
,\nu)=\frac{\delta(\mu(\nu_{1}+\nu_{2})-\nu(\mu_{1}+\mu_{2}))}{4\pi}
$$
\begin{equation}\label{calculatekernel}
\times\exp
\left(\frac{i}{2}\left\{(\nu_{1}\mu_{2}-\nu_{2}\mu_{1})+2X_{1}+2X_{2}-
\left[\frac{\nu_{1}+\nu_{2}}{\nu}+ \frac{\mu_{1}+\mu_{2}}{\mu}
\right]X \right\}\right)
\end{equation}
For the multimode case as well for the infinite dimensional
(functional) case the generalization is in principle
straightforward. In the case of both  Weyl symbols and tomographic
symbols, one simply provides an index (either a discrete or a
continuous one) to the involved ingredients $q$, $p$ and $X$,
$\mu$, $\nu$. In the infinite dimensional case the Wigner symbol
and the tomogram of the operator $\hat{A}$ become functionals.
Correspondingly one modifies the integration measures by the
standard procedure.

\section{Radon transform of Wigner function and Fractional Fourier
transform of wave function}

In this section  we consider the relations of a tomographic symbol
with the Radon transform of the Weyl symbol. In order to write the
Wheeler-DeWitt equation in tomographic form we review some
properties of the modified Radon transform of Wigner function
\bib{wigner32}. The Wigner function is the Weyl symbol
 of the von Neumann density matrix \bib{vonneumann22}.

The Wigner function is expressed in terms of density matrix of
the universe in the form ($\hbar=1$)

\begin{equation}\label{formula1}
 {W}(q,p)=\int\rho\left(q+\frac{u}{2},
 q-\frac{u}{2}\right)e^{-ipu}du
\end{equation}

The inverse transform reads

\begin{equation}\label{formula2}
\rho(x,x')= \frac{1}{2\pi}\int { W} \left(\frac{x+x'}{2},p\right)
e^{ip(x-x')} dp.
\end{equation}

The Radon transform of the Wigner function in the modified form is
the integral transform of the form

\begin{equation}\label{formula3}
  {\cal W} (X,\mu, \nu) = \int W(q,p) e^{ik(X-\mu q- \nu p)}\frac{dkdqdp}{(2 \pi)^{2}}
\end{equation}

Here $X$, $\mu$ $\nu$ are real numbers. The Wigner function can be found using the
inverse Radon relation

\begin{equation}\label{1}
  W(q,p)= \frac{1}{2\pi}\int e^{i(X-\mu q - \nu p)}{\cal
  W}(X,\mu,\nu)dXd\mu d\nu.
\end{equation}
The standard Radon transform is obtained from the two above by
taking $\mu=\cos\varphi$, $\nu=\sin\varphi$.

One can see that the tomographic symbol of density matrix is
given as a marginal distribution since

\begin{equation}\label{3}
  {\cal W}(X,\mu,\nu)= \int W(q,p)\delta(X-\mu q - \nu p)\frac{dqdp}{2\pi}
\end{equation}

It is clear that

\begin{equation}\label{4}
 \int{\cal W} (X,\mu,\nu)dX=1,
\end{equation}
since the Wigner function is normalized

\begin{equation}\label{5}
\int W(q,p)\frac{d q d p}{2\pi}=1
\end{equation}
for normalized wave functions.

The formulae (\ref{3}) -- (\ref{5}) are valid for arbitrary
density matrices, both for pure and mixed states. For pure states
of the universe, the tomographic symbol can be expressed directly
in terms of the wave function of the universe using
\bib{mendespl},

\begin{equation}\label{univwavfun}
 {\cal W}(X,\mu,\nu)=\frac{1}{2\pi|\nu|}\left|\int\psi(y)e^{\frac{i\mu}{2\nu}y^{2}-
 \frac{iX}{\nu}y}dy\right|^{2}.
\end{equation}
The inverse transform provides the wave function due to the
relation
\begin{equation}\label{inverse}
  \psi(y)\psi^{*}(y')= \frac{1}{2\pi}\int{\cal W} (X,\mu,y-y')
  e^{i\left( X-\mu\frac{y+y'}{2}\right)}dXd\mu
\end{equation}
which for the mixed states reads

\begin{equation}\label{mixed}
  \rho(x,x')=\frac{1}{2\pi}\int{\cal W}(X,\mu,x-x')
  e^{i\left( X-\mu\frac{x+x'}{2}\right)}dXd\mu.
\end{equation}

In fact, both the Weyl symbol and the tomographic symbol of
density matrix can be cast into framework of the theory of the
maps of operators acting in Hilbert space of states onto
functions, the pointwise product of functions being replaced by
the star product\bib{marmops}.

The formula relating the tomographic symbol with the wave
function contains the integral

\begin{equation}\label{integral}
 I=\left| \int\psi(y) e^{\frac{i\mu}{2\nu}y^{2}-\frac{iX}{\nu}y}dy\right|
\end{equation}

In case of $\mu=0$, $\nu=1$ this integral is a conventional
Fourier transform of the wave function. For generic $\mu$, $\nu$
the integral is identical to the modulus of Fractional Fourier
transform  of the wave function \bib{mendespl}\bib{margarita}.
Thus, the Radon transform of the Wigner function in the case of
pure states is related to the Fractional Fourier transform of the
wave function. From the linear integral relations for the density
matrix $\rho(x,x')$ and the tomographic probabilities ${\cal
W}(X,\mu,\nu)$ follow the identities (see e.g.
\bib{lecture})

$$\rho(x,x')\to {\cal
W}(X,\mu,\nu),$$

$$ x\to - \left( \frac{\partial}{\partial
X}\right)^{-1}\frac{\partial}{\partial\mu} +
\frac{i}{2}\nu\frac{\partial}{\partial X},$$

\begin{equation}\label{transtom1}
  x'\to - \left( \frac{\partial}{\partial
X}\right)^{-1}\frac{\partial}{\partial\mu} -
\frac{i}{2}\nu\frac{\partial}{\partial X};
\end{equation}

$$ \frac{\partial}{\partial x}\to
\frac{1}{2}\mu\,\frac{\partial}{\partial X} - i\left(
\frac{\partial}{\partial
X}\right)^{-1}\frac{\partial}{\partial\nu},$$

\begin{equation}\label{transtom2}
  \frac{\partial}{\partial x'}\to
\frac{1}{2}\mu\frac{\partial}{\partial X} + i\left(
\frac{\partial}{\partial
X}\right)^{-1}\frac{\partial}{\partial\nu}.
\end{equation}

The physical meaning of the random variable $X$ and the two real
parameters $\mu$ and $\nu$ is the following one \bib{mancinipl}
\bib{marmops}. The variable $X$  is the position of a quantum
particle. But this position is considered in the specific rotated
and scaled reference frame  in  phase-space. The reference frame
is labeled by two parameters $\mu=\exp(\lambda)\cos\theta$,
$\nu=\exp(\lambda)\sin\theta$. The angle $\theta$ is the rotation
angle and  $\lambda$ is the scaling factor. One has to point out
that the tomographic map can be applied to arbitrary functions
which satisfy equations of different types, like elliptic-type and
like wave equation of Klein-Gordon type, see e.g. \bib{marg}

\section{Functional Radon transform and the Wheeler-DeWitt equation}

In order to write the Wheeler-DeWitt generic equation for the wave
functional of the universe, one needs to present the
generalization of the formulae of the previous sections to the
case of functionals which we identify with the functions of an
infinite number of variables $\psi(x_{1},x_{2},\ldots)$. We can
write these functions in the form $\psi(x(k))$. Replacing  $k\to
\tau$ one sees that the functional depends on the function
$\psi(x(\tau))$ where $\tau$ is a continuous variable. One can
extend the notion of functional considering the parameter $\tau$
to be a vector with several continuous components (e.g.,
$\tau=(\tau_{1}, \tau_{2}, \tau_{3}, \tau_{0})$, like space-time
variables). Also the number of functions can be extended such that

 $$x(\tau)\to(x_1(\tau),
\ldots x_k(\tau),\ldots x_N(\tau)).$$

The index $1,2,\ldots N$ can be considered as counting some set of
indices like $(a,b,c,d)\equiv k$ where the numbers $a,b,c,d$ are
tensor indices. In this sense we omit all the indices and will
treat the functional $\psi(x(\tau))$ in the discussed generic
sense, considering $x$ as a vector and $\tau$ as a vector. One
knows that for the notions of derivative for functionals
$\delta\psi(x(\tau))/\delta (x(\tau'))$ can be introduced simply
as a generalization of the partial derivative of a function of
several variables $\partial\psi(x_{i})/\partial x_{k}$. Given an
equation for the density matrix functional one can get the
corresponding equations for the tomographic probability
functional. To this aim one has to use the  replacements
(\ref{transtom1})  and (\ref{transtom2}), modified for an infinite
number of variables, in the equation for the density matrix.

Thus if one has the wave functional $\psi(x(\tau))$, the
corresponding density matrix functional is
\begin{equation}\label{densmatr}
\rho(x(\tau),x'(\tau'))=\psi(x(\tau))\psi^{*}(x'(\tau)).
\end{equation}
One can introduce the Wigner function of the universe defining it
for a pure state as

$$
  W(q(\tau),(p(\tau))=\int\psi\left(q(\tau)+\frac{u(\tau)}{2}\right)
  \psi^{*}\left(q(\tau)-\frac{u(\tau)}{2}\right)
  $$
 \begin{equation}\label{wign}
  \times\, e^{-i\int u(\tau)p(\tau)d\tau}
  {\cal D}[u(\tau)]
\end{equation}
where $ {\cal D}[u(\tau)]$ is the measure in the Fourier
functional integral. The tomographic probability becomes also the
functional ${\cal W}(X(\tau), \mu(\tau),\nu(\tau))$ which is given
in terms of the Wigner functional of the universe as
$${\cal W}(X(\tau), \mu(\tau),\nu(\tau))= \int W\left(q(\tau), p(\tau)\right) \delta [X(\tau)-
  \mu(\tau)q(\tau) -\nu(\tau)q(\tau)]$$
\begin{equation}\label{tomwig}
 \times{\cal D}(q(\tau), p(\tau))
\end{equation}

Thus the tomographic probability functional is given by the above
formula which is the functional Radon transform of the Wigner
functional.

The universe in a model of quantum cosmology is described by a
wave functional which depends on the spatial metric. This wave
functional obeys the Wheeler-DeWitt equation of the form
\bib{whdw}
\begin{equation}\label{wdw}
  \left[ -G_{ijkl}\frac{\delta^{2}}{\delta h_{ij}\delta_{kl}} -\ ^{3}R(h)h^{1/2}+2\Lambda h^{1/2}\right]
  \Psi[h_{ij}]=0
\end{equation}
where $h_{ij}$ is the spatial metric, $G_{ijkl}$ is the metric on
the space of three geometries (superspace)
\begin{equation}\label{metricss}
 G_{ijkl}=\frac{1}{2}h^{-1/2}(h_{ik}h_{jl}+h_{il}h_{jk}-h_{ij}h_{kl})
\end{equation}
and $^{3}R(h) $ is the scalar curvature of the intrinsic geometry
of the three-surface, $\Lambda$ is the cosmological constant. It
means that the density matrix functional and the analog of the
Wigner function in the form of a functional can be introduced as
well as a tomogram functional of the the quantum state of the
universe. Below in Appendix we will write this equation in
tomographic form equation. But to make  transition to tomographic
representation clearer we discuss first simple cosmological
models.

In the following we shall consider different examples of a
homogeneous and isotropic universe. Even if they can be referred
to the same geometry, these  examples are not equivalent from the
quantization point of view. As a matter of fact, it is well-known
that the canonical formulation of gravity leads to the breaking of
the covariance of the theory with respect of the group of four
dimensional diffeomorphisms. Therefore any change of coordinates
does not necessarily lead to a canonical transformation in the
Hamiltonian formulation of General Relativity.

The evolution of the spatial metric is considered in the context
of the space of (spatial) metrics, i.e. the superspace. When a
homogeneous  model is considered, the spatial metric is
parameterized by functions of time and a model equivalent to a
classical particle results. In this case the evolution is
considered in a restricted version of the superspace, i.e. the
so-called minisuperspace. In the case of a
Friedmann-Lemaitre-Robertson-Walker the minisuperspace is a
described by  particle in one dimension. The presence of fields
like a scalar field would eventually extend the minisuperspace
dimensions.

There exist several elaborated examples of minisuperspaces used in
quantum cosmology, below we consider some of these examples.

\section{Some examples of the Wheeler-DeWitt equations}

\subsection{Homogeneous and isotropic universe with cosmological constant and no material source}
In our first example we consider the model in which the metric
dependence is reduced to dependence only on the expansion factor
of the universe. This is a one dimensional Wheeler-DeWitt equation
for a FLRW universe of the form

\begin{equation}\label{flrw}
   \frac{1}{2}\left\{ \frac{1}{a^{p}}\frac{d}{d a}a^{p}\frac{d}{d
   a}-a^{2} +\Lambda a^{4}\right\}\psi(a)=0
\end{equation}

Here $0\leq a <+\infty$, is in the classical theory the expansion
factor and $p$ is an index introduced to take into account the
ambiguity of operator ordering. The Radon transform discussed in
previous sections makes sense only for variables that take values
from $-\infty$ to $+\infty$, so we make the change of variables
$a=\exp{x}$ and the Wheeler-DeWitt equation becomes
\begin{equation}\label{tflrw}
   \frac{1}{2}\left\{  \exp(-2x)\frac{d^{2}}{d x^{2}}+ (p-1)\exp(-2x)\frac{d}{d x}
   -2U(x)\right\}\Psi(x)=0
\end{equation}
where $U(x)=(\exp(2x)-\Lambda\exp(4x))/2$. This equation can be
written also in the form
\begin{equation}\label{tflrwstar}
   \frac{1}{2}\left\{  \exp(-2x')\frac{d^{2}}{d x'^{2}}+ (p-1)\exp(-2x')\frac{d}{d x'}
   -2U(x')\right\}\Psi^{*}(x')=0.
\end{equation}

Multiplying the two equations respectively by $\Psi^{*}(x')$ and
by $\Psi(x)$, and taking the difference, we finally obtain the
equation for the density matrix $\rho(x,x')=\Psi(x)\Psi^{*}(x')$

$$ \frac{1}{2}\Bigg\{  \left[\exp(-2x)\left(\frac{d}{d
x}\right)^{2}- \exp(-2x')\left(\frac{d }{d x'}\right)^{2}\right]
$$
\begin{equation}\label{flrwrho}
+\frac{1}{2}(p-1)\left[\exp(-2x)\frac{d}{d x}-\exp(-2x')\frac{d}{d
x'}\right]
   -(U(x)-U(x'))\Bigg\}\rho(x,x')=0
\end{equation}

Using equations (\ref{transtom1}) and (\ref{transtom2}) we get the
equation

$$\Bigg\{ \mbox{Im}\,\Bigg[\exp\Bigg[2\left( \frac{\partial}{\partial
X}\right)^{-1}\frac{\partial}{\partial\mu}
+i\nu\frac{\partial}{\partial
X}\Bigg]\Bigg(\frac{1}{2}\mu\frac{\partial}{\partial X} - i\left(
\frac{\partial}{\partial
X}\right)^{-1}\frac{\partial}{\partial\nu} \Bigg)^{2}\Bigg]$$

$$+(p-1)\mbox{Im}\Bigg[\exp\Bigg(2\left( \frac{\partial}{\partial
X}\right)^{-1}\frac{\partial}{\partial\mu}
+i\nu\frac{\partial}{\partial
X}\Bigg)\Bigg(\frac{1}{2}\mu\frac{\partial}{\partial X} - i\left(
\frac{\partial}{\partial
X}\right)^{-1}\frac{\partial}{\partial\nu} \Bigg)\Bigg]$$

$$
  - 2\mbox{Im}\Bigg[ \exp\Bigg( - 2\left( \frac{\partial}{\partial
X}\right)^{-1}\frac{\partial}{\partial\mu} +
 i\nu\frac{\partial}{\partial X}\Bigg)-
  \Lambda\exp\Bigg( - 4\left( \frac{\partial}{\partial
X}\right)^{-1}\frac{\partial}{\partial\mu}
 $$
 \begin{equation}\label{tomo}
 + 2i\nu\frac{\partial}{\partial
 X}\Bigg)\Bigg]\Bigg\}\; \mathcal{W}(X,\mu,\nu)=0,
\end{equation}

this equation is the tomographic form of the equation
(\ref{tflrw}). There is no exact solution of equation
(\ref{tflrw}), but for very large $a$ the solution has the form
(see
\bib{hawking2})
\begin{equation}\label{soluzappross}
\psi(a)\sim \cos \frac{Ha^{3}}{3}
\end{equation}
The expression for the tomogram in this case is
\begin{equation}\label{tomo1}
  \mathcal{W}(X,\mu,\nu)=\frac{1}{2\pi|\nu|}\left| \int \cos\frac{Hy^{3}}{3} e^{i\mu y^{2}/2\nu}
  e^{-iXy/\nu}dy\right|^{2}.
\end{equation}

\subsection{Homogeneous and isotropic universe with a different metric}
In \bib{hl}\bib{jl}  a (closed)  homogeneous and isotropic
universe is considered, but where the metric is expressed in a
coordinate system such that it takes the form

\begin{equation}\label{different}
    ds^{2}=-\frac{N^{2}(t)}{q(t)}dt^{2}+q(t)d\Omega_{3}^{2}.
\end{equation}
In this case the Wheeler-DeWitt equation assumes the form
\begin{equation}\label{different2}
    \frac{1}{2}\left( 4\frac{d^{2}}{dq^{2}}+\lambda q
    -1\right)\psi(q)=0,
\end{equation}
where $\lambda$ is a parameter related to  the cosmological
constant $\Lambda$ and the gravitational constant $G$ by the
relation $\lambda=2\; G\; \Lambda /9 \pi$ (see \bib{jl}).

This equation can be expressed in the following form
\begin{equation}\label{airyeq}
\frac{d^{2}\psi(\xi)}{d\xi^{2}}+\xi\psi(\xi)=0
\end{equation}
where
\begin{equation}\label{xi}
    \xi=\left(\frac{\lambda}{2}\right)^{1/3}\left(q-\frac{1}{\lambda}\right).
\end{equation}
The solution of equation (\ref{airyeq}) is
\begin{equation}\label{solution}
   \psi(q)=A\Phi(-\xi)=A\Phi \left(-\frac{\lambda}{2}\right)^{1/3}\left(q-\frac{1}{\lambda}\right)
\end{equation}
where $\Phi(x)$ is the Airy function
\begin{equation}\label{airyfunc}
\Phi(x)=\int_{0}^{\infty}\cos\left(\frac{u^{3}}{3}+u\,x\right)du,
\end{equation}
$A$ is a normalization constant.
 The equation for the tomogram is the same equation for the tomogram of an electric
 charge moving in a constant homogeneous electric field and it reads \bib{shchukin}
\begin{equation}\label{tomohomogisotro}
    -\mu \frac{\partial \mathcal{W}}{\partial \nu}+ F \nu \frac{\partial \mathcal{W}}{\partial
    X}=0
\end{equation}
The corresponding expression for the tomogram is
\begin{equation}\label{tomoairy}
  \mathcal{W}(X,\mu,\nu)=\frac{A^{2}}{2\pi|\nu|}\left| \Phi\left( -\left(\frac{1}{2\lambda^{2}}\right)^{1/3}
+ \left(\frac{\lambda}{2}\right)^{1/3}\frac{X}{\mu}
-\left(\frac{\lambda}{2}\right)^{2/3} \frac{\nu^{2}}{\mu^{2}}
\right) \right|^{2}
\end{equation}

\subsection{The ``harmonic oscillator'' case}
 Hartle and Hawking showed \bib{hawking2} that the Wheeler-DeWitt equation for a
homogeneous and isotropic metric

\begin{equation}\label{homisomet}
    ds^{2}=\sigma^{2}(N^{2}d\tau^{2}+a^{2}(\tau)d\Omega_{3}^{2}),
\end{equation}
 with a conformally invariant
 field $\varphi$ and zero cosmological constant, reduces to the
equation of a harmonic oscillator
\begin{equation}\label{harmoscill}
   \frac{ 1}{2}\left( \frac{\partial^{2}}{\partial x^{2}} - \omega_{1}^{2}x^{2} -
   \frac{\partial^{2}}{\partial y^{2}} +\omega_{2}^{2}y^{2} \right)
   \psi(x,a)=0
\end{equation}
with $x= a$, $y=\phi\,a$ and $\omega_{1}=\, \omega_{2}=1$. But
Gousheh and Sepangi \bib{gousheh} pointed out that the  equation
(\ref{harmoscill}) holds also for other cosmological models. For
example by taking a scalar field $\phi$ with potential
\begin{equation}\label{potential}
    V(\phi)=\lambda+\frac{m^{2}}{2\alpha^{2}}\sinh^{2}(\alpha
    \phi)+\frac{b}{2\alpha^{2}}\sinh(2\alpha
    \phi)
\end{equation}
one can obtain, by suitable changes of coordinates, equation
(\ref{harmoscill}). The same equation can be derived in a
Kaluza-Klein cosmology with negative cosmological constant and
metric
\begin{equation}\label{kalklemetric}
   ds^{2}=-dt^{2}
   +a^{2}(t)\frac{\delta_{ij}dx^{i}dx^{j}}{(1+\frac{kr^2}{4})}+A^{2}(t)d\varrho^{2}
\end{equation}
where $A(t)$ is the radius of the compactified dimension.

A solution of equation (\ref{harmoscill}) can be obtained by
separation of variables \bib{gousheh}

\begin{equation}\label{solutionharosc}
\psi_{n_{1}n_{2}}(x,y)=\alpha_{n_{1}} (x) \beta_{n_{2}}(y)\ \ \ \
n_{1} , n_{2}=\, 0, 1, 2, \dots
\end{equation}
where both the families of functions $\alpha_{n}(x)$ and
$\beta_{n}(y)$ are expressed by
\begin{equation}\label{alfaconenne}
     \alpha_{n}(x)=\left(\frac{1}{\pi}\right)^{1/4}
     \frac{H_{n}( x)}{\sqrt{2^{n}n!}}e^{-\,x^{2}/\,2}
\end{equation}
and
\begin{equation}\label{betaconenne}
     \beta_{n}(y)=\left(\frac{1}{\pi}\right)^{1/4}
     \frac{H_{n}(y)}{\sqrt{2^{n}n!}}e^{-\,y^{2}/\,2}.
\end{equation}
One can obtain by the described method the corresponding equation
for the tomogram and it reads
\begin{equation}\label{tomharoscil}
    -\mu_{1}\frac{\partial \mathcal{W}}{\partial \nu_{1}}+\nu_{1}\frac{\partial \mathcal{W}}{\partial
    \mu_{1}}+\mu_{2}\frac{\partial \mathcal{W}}{\partial \nu_{2}}-
    \nu_{2}\frac{\partial \mathcal{W}}{\partial
    \mu_{2}}=0
\end{equation}

The corresponding solution can be found  by applying equation
(\ref{univwavfun}) to  (\ref{solutionharosc}) and we obtain the
tomographic symbol
$$
{\mathcal
     W}_{n_{1}n_{2}}(X_{1},\mu_{1},\nu_{1},X_{2},\mu_{2},\nu_{2})
$$

$$
=\frac{1}{(2\pi)^{2}|\nu_{1}{\nu_{2}}|}\left|
\int\psi_{n_{1}n_{2}}(x,y)e^{\frac{i\mu_{1}x^{2}}{2\nu_{1}}}e^{\frac{i\mu_{2}y^{2}}{2\nu_{2}}}
e^{-i\frac{x X_{1}}{\nu_{1}}}e^{-i\frac{y X_{2}}{\nu_{2}}}dx\, dy
\right|^{2}
$$

\begin{equation}\label{tomsymb}
 =\frac{1}{\pi}\frac{1}{2^{n_{1}+n_{2}}}\frac{1}{n_{1}!n_{2}!}
\frac{e^{\,\frac{-X_{1}^{2}}{\mu_{1}^{2}+\nu_{1}^{2}}} \;
e^{\,\frac{-X_{2}^{2}}{\mu_{2}^{2}+\nu_{2}^{2}}}}
{\sqrt{(\mu_{1}^{2}+\nu_{1}^{2})(\mu_{2}^{2}+\nu_{2}^{2})}}\,
H^{2}_{n_{1}}\left(\frac{X_{1}}{\sqrt{(\mu_{1}^{2}+\nu_{1}^{2})}}\right)
H^{2}_{n_{2}}\left(\frac{X_{2}}{\sqrt{(\mu_{2}^{2}+\nu_{2}^{2})}}\right).
\end{equation}

With the solutions found above, we can describe the tomogram for
an entangled state of the universe. Entangled systems were already
considered in the context of General Relativity by Basini et al.
\bib{capozz}. For instance, let us consider the combination which
is the entangled superposition state of the universe in the model
under study
\begin{equation}\label{combination}
    \frac{1}{\sqrt{2}}\psi_{12}+\psi_{21}=
    \frac{1}{\sqrt{\pi}}(ye^{-\frac{x^{2}}{2}}e^{-\frac{y^{2}}{2}}+xe^{-\frac{x^{2}}{2}}e^{-\frac{y^{2}}{2}});
\end{equation}
the corresponding tomogram is
$$
{\mathcal
     W}_{12}^{\,{\rm entangled}}(X_{1},\mu_{1},\nu_{1},X_{2},\mu_{2},\nu_{2})
$$

$$
=\frac{1}{2(2\pi)^{2}|\nu_{1}{\nu_{2}}|}\left|
\int(\psi_{12}(x,y)+\psi_{21}(x,y))e^{\frac{i\mu_{1}x^{2}}{2\nu_{1}}}e^{\frac{i\mu_{2}y^{2}}{2\nu_{2}}}
e^{-i\frac{x X_{1}}{\nu_{1}}}e^{-i\frac{y X_{2}}{\nu_{2}}}
\right|^{2}
$$

$$
=\frac{1}{(2\pi)^{2}|\nu_{1}{\nu_{2}}|}\left|
  2\sqrt{\pi} \left(\frac{X_{1}(\mu_{1}-i\nu_{1}))}{(\mu_{1}^{2}+\nu_{1}^{2})}
 +\frac{X_{2}(\mu_{2}-i\nu_{2}))}{(\mu_{2}^{2}+\nu_{2}^{2})}
 \right)\right.
 $$
\begin{equation}\label{entangleduniverse}
 \left.\times\sqrt{\frac{1+i\mu_{1}/\nu_{1}}{1+\mu^{2}_{1}/\nu^{2}_{1}}\cdot\frac{1+i\mu_{2}/\nu_{2}}{1+\mu^{2}_{2}/\nu^{2}_{2}}}
\;
e^{-\frac{X_{1}^{2}(\nu_{1}+i\mu_{1})}{2(\mu_{1}^{2}+\nu_{1}^{2})}}\;e^{-\frac{X_{2}^{2}(\nu_{2}+i\mu_{2})}{2(\mu_{2}^{2}+\nu_{2}^{2})}}
\right|^{2}
\end{equation}

$$
=\frac{1}{\pi}
 \left(\frac{X_{1}^{2}}{(\mu_{1}^{2}+\nu_{1}^{2})}
 +\frac{X_{2}^{2}}{(\mu_{2}^{2}+\nu_{2}^{2})}+
 \frac{2X_{1}X_{2}(\mu_{1}\mu_{2}+\nu_{1}\nu_{2})}{(\mu_{1}^{2}+\nu_{1}^{2})(\mu_{2}^{2}+\nu_{2}^{2})}
 \right)
 $$
\begin{equation}\label{entangleduniverse2}
 \times\sqrt{\frac{1}{\nu^{2}_{1}+\mu^{2}_{1}}\cdot\frac{1}{\nu^{2}_{2}+\mu^{2}_{2}}}
\;
e^{-\frac{X_{1}^{2}}{(\mu_{1}^{2}+\nu_{1}^{2})}}\;e^{-\frac{X_{2}^{2}}{(\mu_{2}^{2}+\nu_{2}^{2})}}.
\end{equation}

This tomogram is the positive joint probability distribution of
two random variables $X_{1}$ and $X_{2}$ and it completely
determines the quantum state  of the universe in the considered
model.
\section{Conclusions}

To conclude we summarize the main results of our work. In the
framework of quantum gravity we applied the recently introduced in
quantum mechanics and quantum optics method of association with
quantum states the probability distributions and in in view of
this we managed to describe the states of the universe by standard
positive probability distributions (tomograms of the universe
states). We found the connection of this approach with star
product (deformation) quantization.   The conventional
Wheeler-DeWitt for the wave function of the universe is presented
in the form of a stochastic equation for the standard positive
probability distributions. The wave function of the universe and
its density matrix can be reconstructed in terms of the introduced
tomographic probability distribution of the universe. Some example
of Friedmann-Lemaitre-Robertson-Walker minisuperspaces were
explicitly studied and the tomograms of the corresponding universe
states were showed including an entangled state. The description
of  an universe quantum state by standard positive probability
distributions provides some new aspects to the problem of the
connection with the classical description of pure universe states
.

It is worthy to study these new aspects considering the classical
limit  of the quantum equations in the tomographic representation.
One has to point out that for studying the classical limit one
needs to take into account the decoherence phenomena which destroy
the quantum coherence of the universe states. The classical limit
of a quantum mechanical problem (kicked rotators) was discussed in
tomographic representation in
\bib{mendes2}.

Acknowledgements

V.I.M thanks the University ``Federico II'' and INFN sezione di
Napoli for hospitality and Russian Basic Research Foundation
(Grant 01-02-17745) for partial support.

\section{Appendix}
In the Appendix we review the details of the Radon transform
approach to the Schr\"odinger equation and the Von Neumann
equation. To do this we describe how the Schr\"odinger equation
for the wave function induces the Von Neumann equation for the
density matrix. After this the tomographic transform provides the
equation for the tomogram of a quantum state.

The Schr\"odinger evolution equation for a system with one degree
of freedom  reads ($m=1$)
\begin{equation}\label{onedim}
  i\frac{\partial\psi(x,t)}{\partial t}
  =-\frac{1}{2}\frac{\partial^{2}\psi(x,t)}{\partial x^{2}}+
  U(x)\psi(x,t);
\end{equation}
$\hbar=1$.

The Schr\"odinger equation energy-level equation reads
\begin{equation}\label{enelev}
-\frac{1}{2}\frac{\partial^{2}\psi_{E}(x,t)}{\partial x^{2}}+
  U(x)\psi_{E}(x,t)=E\psi_{E}(x,t)
\end{equation}
For the density matrix
\begin{equation}\label{densmatrix}
  \varrho(x,x',t)=\psi(x,t)\psi^{*}(x',t)
\end{equation}
the von Neumann evolution equation can be obtained from equation
(\ref{onedim}) and it has the form
\begin{equation}\label{vonneu}
i\frac{\partial\varrho(x,x',t)}{\partial t}
  =-\frac{1}{2}\left[\frac{\partial^{2}\varrho(x,x',t)}{\partial x^{2}}-
  \frac{\partial^{2}\varrho(x,x',t)}{\partial {x'}^{2}}\right]+
  \left(U(x)-U(x')\right)\varrho(x,x',t)
\end{equation}

Using the relations (\ref{univwavfun}) and (\ref{mixed}) one can
see that the evolution equation for the tomogram of the quantum
state can be obtained from the evolution equation for  the density
matrix (\ref{vonneu}) using the replacements
\begin{equation}\label{replac}
 \varrho(x,x',t)\rightarrow {\mathcal{W}}(X,\mu,\nu,t),
\end{equation}
 (\ref{transtom1}) and (\ref{transtom2}).
 Thus the evolution equation for the quantum state tomogram has
 the form \bib{mancinipl}
\begin{equation}\label{qst}
   \frac{\partial\mathcal{W}}{\partial
   t} -\mu\frac{\partial\mathcal{W}}{\partial\nu}+  \left[U(\tilde
   {q})-U(\tilde{q}^{*})\right]{\mathcal{W}}= 0
\end{equation}
where the argument of the potential is the operator

\begin{equation}\label{operator}
\tilde q= -\left(\frac{\partial}{\partial
X}\right)^{-1}\frac{\partial}{\partial\mu}+ i
\frac{\nu}{2}\frac{\partial}{\partial X}.
\end{equation}
Here the operator $(\partial/\partial X)^{-1}$ is defined by the
action onto the Fourier component $\tilde f(k)$ of a function
$f(x)$
\begin{equation}\label{fourier}
 f(x)= \int\tilde f(k)e^{ikx}dk
\end{equation}
due to the prescription
\begin{equation}\label{prescrip}
  \left(\frac{\partial}{\partial
X}\right)^{-1} f(x)=\int\frac{\tilde f(k)}{ik}e^{ikx}dk
\end{equation}
The evolution equation for the tomogram (\ref{qst}) is the
tomographic map of the Moyal equation \bib{moyal49} for the Wigner
function $W(q,p,t)$

\begin{equation}\label{wqst}
   \frac{\partial {W}}{\partial
   t} +p\frac{\partial {W}}{\partial q}+  \left[U(\tilde{\tilde
   {q}})-U(\tilde{\tilde{q}}^{*})\right] {W}= 0
\end{equation}
where the argument of the potential is the operator
\begin{equation}\label{argument}
   \tilde{\tilde q} = q + \frac{i}{2}\frac{\partial}{\partial p}.
\end{equation}
Thus introducing the functional
\begin{equation}\label{functional}
\rho(x,x')=\psi(x)\psi^{*}(x')
\end{equation}
we get the Wheeler-DeWitt equation for the tomogram of the
universe by means of the replacements
\begin{equation}\label{replace1}
x\to - \left(\frac{\delta}{\delta
X}\right)^{-1}\frac{\delta}{\delta \mu} +
\frac{i}{2}\nu\frac{\delta}{\delta X},
\end{equation}
\begin{equation}\label{replace2}
x'\to - \left(\frac{\delta}{\delta
X}\right)^{-1}\frac{\delta}{\delta \mu} -
\frac{i}{2}\nu\frac{\delta}{\delta X}
\end{equation}
\begin{equation}\label{replace3}
\frac{\delta}{\delta x}\to\frac{1}{2}\mu\frac{\delta}{\delta X} -
i\left(\frac{\delta}{\delta X}\right)^{-1}\frac{\delta}{\delta\nu}
\end{equation}

\begin{equation}\label{replace4}
\frac{\delta}{\delta
x'}\to\frac{1}{2}\mu\frac{\delta}{\delta X} +
i\left(\frac{\delta}{\delta X}\right)^{-1}\frac{\delta}{\delta\nu}
\end{equation}

which should be done in analogy with the von Neumann equation for
the  density of the universe \bib{whdw}
$$
\left[-F_{\alpha\beta}\frac{\delta^{2}}{\delta x_{\alpha}\delta x
_{\beta}}-3R(x)S(x) + 2 \Lambda S(x)\right.
$$
\begin{equation}\label{funzionale}
\left. -F_{\alpha\beta}(x')\frac{\delta^{2}}{\delta
x'_{\alpha}\delta x'_{\beta}}-3R(x')S(x') + 2 \Lambda
S(x')\right]\varrho(x,x')=0
\end{equation}
Here
\begin{equation}\label{F}
F_{\alpha\beta}(x)= - G_{ijkl},
\end{equation}
\begin{equation}\label{S}
 S(x)=h^{\frac{1}{2}},
\end{equation}
\begin{equation}\label{R}
 R(x)=\ ^{3}R(h).
\end{equation}

We take into account that the wave function of the universe is a
real function. Making in (\ref{funzionale}) the replacement
$$
\varrho(x,x')\to \mathcal{W}(X,\mu,\nu)
$$
and using equations (\ref{replace1})--(\ref{replace4}), we get the
tomographic form of the Wheeler-DeWitt equation.

The considered cosmological models are particular cases of this
general procedure.

%
%


\begin{thebibliography}{99}
\item\label{radon} J. Radon, Ber. Verh. Sachs. Acad. {\bf 69}, 269
(1917)
 \item\label{vonneumann22} J. von Neumann ``Mathematische
Grundlagen der Quantenmechanik'', Springer Verlag, Berlin 1932
\item\label{landau} L. D. Landau, Z. Physik, {\bf 45} 430 (1927)
\item\label{wigner32} E. Wigner, Phys. Rev. {\bf 40}, 749 (1932)
\item\label{vogrisken} K. Vogel, H. Risken Phys. Rev {\bf A40},
2847 (1989)
\item\label{raymer}  D. T. Smithey, M. Beck, M. G.
Raymer, A. Faridani  Phys. Rev Lett. {\bf 70}, 1244 (1993)

\item\label{bertr} J.Bertrand, D. Bertrand, Found.Phys. {\bf 17},
397 (1987) \item\label{mancini95} S. Mancini, V. I. Manko, P.
Tombesi, Quant. Semiclass. Opt. {\bf 7}, 615 (1995)
\item\label{mancinipl} S. Mancini, V. I. Manko, P. Tombesi, Phys.
Lett. {\bf A213}, 1, (1996)
 \item\label{marmosud} V.I. Manko, G.~Marmo, E.~C.~G.~Sudarshan,
 F.~Zaccaria, J. Russ. Laser Res. (New York: Plenum Publ.) {\bf
 20}, 421 (1999); Phys. Lett. {\bf A273}, 31 (2000); J. Phys. A:
 Math. and General {\bf 35}, 7137 (2002).

 \item\label{marmops} O. V.
Manko, V.I. Manko, G.~Marmo, Phys. Scr.{\bf62},446,(2000); J.
Phys. A {\bf 35}, 699 (2001)

\item\label{moyal49}  J.~E.~Moyal, Proc. Cambridge Philos. Soc.
{\bf 45}, 99 (1949)
 \item\label{hawking1} S. W. Hawking, Nucl. Phys {\bf B239}, 257 (1984)
 \item\label{whdw} B. S. DeWitt Phys. Rev.  {\bf 160}, 1113, (1983); J. A  Wheeler in
 \textit{Battelle Rencontres}, edited by C. DeWitt and J. A. Wheeler (Benjamin, New
 York, 1968).

 \item\label{hh} J. B. Hartle and S. W. Hawking, Phys. Rev. D {\bf 28}, 2960, (1983)
 \item\label{hawking2} S. W. Hawking, in ``Intersection between Elementary Particle Physics and Cosmology'' eds. T. Piran
  and S. Weinberg, World Scientific Publishing Co (Singapore)(1986)
 \item\label{dnp} D. N. Page Phys. Rev. D {\bf 34}, 2267, (1986)
\item\label{hl} J. J. Halliwell, Phys. Rev. D {\bf 38}, 2468, (1988)

 \item\label{rubakov}V.~A.~Rubakov,``Quantum cosmology,'' arXiv:gr-qc/9910025.
 \item\label{parentani} R. Parentani, Phys. Rev. D {\bf 56}, 4618
 (1997)
 \item\label{anton} F. Antonsen, ``Deformation Quantisation of
 Gravity'', gr-qc/9712012
 \item\label{kodama}H.~Kodama, ``Quantum Cosmology In Terms Of The Wigner Function,''
KUCP-0014 {\it Presented at 5th Marcel Grossmann Mtg., Perth,
Western Australia, Aug 8-12, 1988}


  \item\label{mendespl} V. I. Manko, R.V. Mendes, Phys. Lett.
{A263}, 53 (1999)
\item\label{asht} A. Ashtekar, ``Lectures on Non-perturbative
canonical gravity'' World Scientific (1991); A. Ashtekar, ``New
perspectives in canonical gravity'' Bibliopolis (Napoli) (1988)

\item\label{baez} J.C. Baez, , ``Diffeomorphism-invariant generalized measures on
the space of connections modulo gauge transformations'', in Crane,
L., and Yetter, D., eds., Proceedings of the Conference on Quantum
Topology, 213-223, (World Scientific, Singapore, 1994).
 \item\label{rosapl} V. I. Manko, L. Rosa, P. Vitale
Phys. Lett B {\bf 439}, 328 (1998)
 \item\label{fronsdal} F. Bajen,
M. Flato, M. Fronsdal, C. Lichnerowicz, D.
 Sternheimer, Lett. Math. Phys. {\bf 1}, 521 (1975)



 \item\label{margarita}M. A. Manko, J. Russ. Laser Res. {\bf 21},
 421 (2000)


 \item\label{lecture}O. V. Manko, V. I. Manko, J. Russ. Laser Res. {\bf 18},
 407 (1997)
\item\label{marg}M. A. Manko, J. Russ. Laser Res. {\bf 22},
 168 (2001)
\item\label{jl} J. Louko Class. Quantum Grav. \textbf{4},  581, (1987)
\item\label{shchukin} V. I. Manko, E. V. Shchukin J. Russ. Laser
Res. {\bf 22}, 545 (2001)

 \item\label{gousheh} S. S. Gousheh, H. R. Sepangi, Phys. Lett.
{A272}, 304 (2000)

 \item\label{capozz}G.~Basini, S.~Capozziello and G.~Longo,
 Astron. Nach. {\bf 324}, 275 (2003); La Rivista del N. Cim. {\bf
 25}, N.11 (2002)

 \item\label{mendes2} V.I. Manko, R. V. Mendes,
Physica {\bf D145}, 330 (2000)




\end{thebibliography}
\end{document}